\documentstyle[12pt]{article}
\renewcommand{\thefootnote}{\fnsymbol{footnote}}
\begin{document}
\begin{flushright}
Columbia preprint CU--TP--695
\end{flushright}
\vspace*{1cm}
\setcounter{footnote}{1}
\begin{center}
{\Large\bf The Phase Transition to the Quark--Gluon Plasma
and Its Effect on Hydrodynamic Flow\footnote{This work was supported by the
Director, Office of Energy Research, Division of Nuclear Physics of
the Office of High Energy and Nuclear Physics of the U.S. Department
of Energy under Contract No.\ DE-FG-02-93ER-40764.}}
\\[1cm]
Dirk H.\ Rischke\footnote{Partially supported by the
Alexander von Humboldt--Stiftung under
the Feodor--Lynen program.} \\ ~~ \\
{\small Physics Department, Pupin Physics Laboratories, Columbia
University} \\
{\small 538 W 120th Street, New York, NY 10027, U.S.A.} \\ ~~ \\
Yar\i\c{s} P\"urs\"un, Joachim A.\ Maruhn, Horst St\"ocker,
Walter Greiner \\ ~~ \\
{\small Institut f\"ur Theoretische Physik der J.W.\ Goethe--Universit\"at} \\
{\small Robert--Mayer--Str.\ 10, D--60054 Frankfurt/M., Germany}
\\ ~~ \\ ~~ \\
{\large May 1995}
\\[1cm]
\end{center}
\begin{abstract}
It is shown that in ideal relativistic hydrodynamics
a phase transition from hadron to quark and gluon degrees
of freedom in the nuclear matter equation of state
leads to a minimum in the excitation function of the
transverse collective flow.
\end{abstract}
\renewcommand{\thefootnote}{\arabic{footnote}}
\setcounter{footnote}{0}
\newpage
Hydrodynamics represents (local) energy, momentum, and charge
conservation \cite{maruhn,stoecker,strottman}. Because of its simplicity
it has found widespread application in studying the dynamical evolution
of heavy--ion collisions (see e.g.\ \cite{stoecker,strottman,st2} and refs.\
therein).
For instance, 3+1--dimensional hydrodynamical calculations for
collisions at BEVALAC energies ($E_{Lab}^{kin}= 0.1 - 2$ AGeV) were performed
about twenty years ago (see the very detailed review of
this topic in \cite{stoecker}).
It was found that the compressional shock waves created in the collision
lead to collective flow phenomena like sideward deflection
of matter in the reaction plane (``side-splash'' and ``bounce-off'')
as well as azimuthal deflection out of the reaction plane (``squeeze-out'').
The confirmation of these collective flow effects by BEVALAC experiments
\cite{gutbrod,rai} was one of the main successes of the
fluid-dynamical picture.

Inspired by this success, 3+1--dimensional ideal relativistic hydrodynamics
was also applied to study heavy--ion collisions at AGS ($E_{Lab}^{kin}
\simeq 10 - 15$ AGeV) and even CERN--SPS energies ($E_{Lab}^{kin} \simeq
60 - 200$ AGeV) \cite{st2,rentzsch,csernai,bravina}. Since ideal
hydrodynamics assumes that matter is in local equilibrium at every instant,
colliding fluid elements are forced by
momentum conservation to instantaneously stop
and by energy conservation to convert all their kinetic energy
into internal energy (compression and heating via shock waves).
The longitudinal rapidity loss in
individual nucleon--nucleon collisions is, however, limited. Thus, immediate
complete stopping is not achieved in reality and, for higher beam
energies, it is no longer justified to treat the initial stage of the reaction
in an ideal hydrodynamical picture. Ideal hydrodynamics might nevertheless
be applicable in the expansion stage of the collision \cite{prakash},
where the conditions of local thermodynamical equilibrium
are more likely to be established. In order
to describe the initial stage, however, one has
to account for non-equilibrium effects \cite{strottman}.

These effects are naturally accounted for in
microscopic transport models like QMD \cite{QMD}, RQMD \cite{RQMD},
ARC \cite{ARC}, or ART \cite{ART}. For BEVALAC, SPS, and
AGS energies, these models
explain almost all single particle observables satisfactorily in terms
of hadronic physics (see, for instance, \cite{QMD,QM93} and refs.\ therein).
However, the basic assumption of these models is that the
original (quantum) many--body problem can be adequately decomposed in terms
of (classical) two--particle scatterings. For heavy--ion collisions, this
assumption is up to now unproven.
Moreover, to describe these two--particle scattering events the
measured free hadron--hadron cross sections are employed,
at least as far as these are known experimentally. This leaves
a large uncertainty with respect to unknown cross sections for heavier
mass resonances and effects of the nuclear environment.
Given the fact that microscopic models deal with this uncertainty by
introducing a large number of parameters, it is highly desirable and
of considerable interest and importance to investigate, to what extent
the simpler (and well defined) hydrodynamical theory is applicable
to describe heavy--ion collisions in the beam energy
range from $0.1$ to $10$ AGeV.

The simplicity of the (ideal) hydrodynamical approach lies in the fact that
the only physical input is the nuclear matter equation of state
(EoS) which is calculable by means of thermodynamics \cite{stoecker}.
This constitutes also a main advantage of hydrodynamics over microscopic
models, since e.g.\ the phase transition to the quark--gluon plasma (QGP)
\cite{QGP} can be studied in a simple, straightforward
manner\footnote{Non-equilibrium phenomena like supercooling and bubble
formation \cite{bubble} are neglected in this picture.},
whereas microscopic transport models require an ad hoc
deconfinement--hadronization mechanism \cite{werner}. As we shall see,
the phase transition to the QGP may indeed play a decisive role for
the dynamical evolution of the system in the beam energy range
$E_{Lab}^{kin}=0.1-10$ AGeV.

As discussed in \cite{test1,test2} (see also \cite{bugaev,olli}),
a phase transition in the EoS qualitatively changes the hydrodynamical
flow pattern. In particular, the hydrodynamically stable solution for
one-dimensional, stationary compression of matter is no longer
a single shock wave, but a sequence of shock and compressional simple waves.
Analogously, the hydrodynamically stable solution for one-dimensional,
stationary expansion is a sequence of simple rarefaction waves and
rarefaction shock discontinuities, instead of a single simple rarefaction wave.

In \cite{DHRMG} it is shown (for matter without a
conserved charge) that these phenomena occur not only
if the EoS has a first order phase transition, but also
if the energy (or entropy) density rises sufficiently
rapidly as a function of the temperature.
But even if this increase is only moderate, and a simple rarefaction
wave is the hydrodynamically stable expansion solution, the flow pattern
shows structures resembling broadened (``smeared'') versions of
rarefaction shocks. Decisive for their occurrence
is the existence of a so-called ``softest point'' in the EoS \cite{hung},
i.e., a local minimum of $p/\epsilon$ as a function of $\epsilon$ ($p$
is the pressure, $\epsilon$ the energy density in the local rest frame
of matter).

As pointed out in \cite{test1,hung}, the existence of this ``softest
point'' leads to a prolonged expansion of matter and consequently to a
long lifetime of a mixed phase of QGP and hadron matter.
Analogously, it also takes longer to compress matter in the early stage
of a heavy--ion collision \cite{test2}.
In this work we shall demonstrate in detail how both effects lead to a
{\em minimum in the excitation function of the directed
transverse collective flow in heavy--ion collisions}. An observation of this
phenomenon would be a clear signature for a change of nuclear matter properties
as for instance in the transition from hadron to quark and gluon degrees
of freedom. We remark that our results are in agreement with earlier
works \cite{csernai,bravina} (see also \cite{hofmann})
which indicated a decrease of the transverse collective
flow if the nuclear matter EoS features a phase transition to the QGP.

The relativistic hydrodynamical equations represent
(local) energy--momentum conservation
\begin{equation} \label{dt}
\partial_{\mu} T^{\mu \nu} =0
\end{equation}
and (local) charge conservation
\begin{equation} \label{dn}
\partial_{\mu} N^{\mu} = 0~.
\end{equation}
Here $T^{\mu \nu}$ is the energy--momentum tensor and $N^{\mu}$ the
net baryon number current.
In case that also other conserved quantum numbers are to be considered
(like e.g.\ net strangeness) there is an additional equation of
the type (\ref{dn}) for each of these charges.
Provided that matter is in local thermodynamical equilibrium, the
energy--momentum tensor $T^{\mu \nu}$ and the baryon current $N^{\mu}$
assume ideal fluid form \cite{LL6}, i.e.,
\begin{eqnarray} \label{t}
T^{\mu \nu} & = & (\epsilon +p) u^{\mu} u^{\nu} - p g^{\mu \nu}~,\\
N^{\mu} & = & n u^{\mu}~, \label{n}
\end{eqnarray}
where $n$ is the baryon number density in the local rest frame of the fluid,
$u^{\mu} \equiv \gamma
(1,{\bf v})$ is the fluid 4--velocity ($\gamma \equiv (1-{\bf v}^2)^{-1/2}$,
${\bf v}$ is the fluid 3--velocity), and $g^{\mu \nu} = {\rm diag}(+,-,-,-)$
is the metric tensor.
The equations of ideal fluid-dynamics are closed by specifying the
nuclear matter EoS in the form $p=p(\epsilon,n)$.

This EoS is constructed as follows (for
details see \cite{test2}). For the QGP phase, the MIT bag EoS \cite{MIT}
(for massless gluons and $u$ and $d$ quarks) is employed (with
a Bag constant $B=(235\, {\rm MeV})^4$), while
the hadronic phase is described by a version of the
$\sigma-\omega$--model \cite{migdhr} (plus massive thermal pions)
which features more realistic values for the ground state incompressibility
($K_0 \simeq 300$ MeV)
and the effective nucleon mass ($M_0^* \simeq 0.635\, M$, where
$M$ is the free nucleon mass) than the original version proposed
by Walecka (where $K_0 \simeq 550$ MeV, $M_0^* \simeq 0.54\, M$)
\cite{walecka}. Both equations of state are matched via
Gibbs' conditions of phase equilibrium. Thus, the resulting nuclear matter
EoS has by construction a first order phase transition between hadron
and quark--gluon matter. In Fig.\ 1 we show the pressure
as a function of energy density and baryon number density.
In the QGP phase the pressure is independent of $n$, since
for the bag model one has the simple relationship $p=(\epsilon-4B)/3$.
The mixed phase is distinct from the other phases in
that the pressure is only slowly varying with $\epsilon$ and $n$.
The hadronic phase corresponds to
the small strip between mixed phase and unphysical $(\epsilon,n)$--combinations
where the pressure is zero\footnote{Note that an ideal gas of nucleons
has a minimum energy density for a given baryon density, namely the Fermi
energy density at $T=0$.}.
In the following we will compare results obtained with this EoS to those
obtained with an EoS for {\em pure\/} hadronic matter, i.e., matter described
{\em solely\/} by the above mentioned hadron matter EoS.

Let us first investigate compression of nuclear matter in a simple
{\em one-dimensional\/} ``slab-on-slab'' collision. The hydrodynamical
solution to this problem was discussed in detail in \cite{test2}. In Fig.\ 2 we
present various thermodynamic quantities in the final compressed state
as functions of the beam energy (of a fixed--target experiment) for
the above described EoS with phase transition (full line) in comparison to
quantities obtained in a single shock compression of pure hadronic matter
(dashed line).
One clearly observes in Fig.\ 2(c) that in the case of a phase transition
 the pressure does not increase as fast with the beam energy
as in the pure hadron matter case. The reason is that in the mixed phase
the pressure increases only slowly as a function of energy density and baryon
number density, cf.\ Fig.\ 1. This leads to the existence of a
``softest point'' at the phase boundary between mixed and quark--gluon
matter, cf.\ Fig.\ 2(e), corresponding to $E_{Lab}^{kin} \simeq 4.1$ AGeV.
As a consequence, for the same beam energy matter
is much easier to compress, i.e., larger energy
and baryon number densities can be obtained than for the ``stiff''
pure hadron matter EoS, cf.\ Figs.\ 2(a,b).
Note that the specific entropy $\sigma \equiv s/n$ in Fig.\ 2(d) is
(approximately) constant as a function of the beam energy in the mixed phase.
Assuming subsequent adiabatic ($\sigma=const.$) expansion of matter until
freeze-out, it was shown in \cite{BGR} that a corresponding plateau
should occur in the excitation function of the pion multiplicity. This
plateau could serve as a signature for QGP formation.

We remark that for the above constructed EoS the phase transition sets in
at $E_{Lab}^{kin} \simeq 1.5$ AGeV and pure QGP is formed
at $E_{Lab}^{kin} \simeq 4.1$ AGeV. These
values appear rather low. The reason is that the hadronic part of
our EoS features only nucleons and pions.
As a consequence, the pressure as a function
of temperature and chemical potential is still quite small at
the phase transition. If more hadronic resonances are included, to first
approximation (i.e., assuming them to behave as ideal gases) their
partial pressures add to the total pressure. This shifts the phase transition
to larger values of temperature, chemical potential, and consequently
to higher energy and baryon number densities. This will in turn also
shift the onset of the phase transition to higher beam energies ($\sim
10$ AGeV according to the results of \cite{RiFri}). However, apart from
such a shift the results of Fig.\ 2 should remain qualitatively
unchanged.

Fig.\ 3 shows the time (in the CM frame of two equal nuclei)
the compression waves need to completely compress
the incoming nuclei \cite{test2}.
One observes that this time is prolonged if the EoS features a phase
transition (full line) as compared to the case where pure hadronic
matter is compressed (dotted line).
The reason is intuitively clear, since a higher compression
reduces the velocity of
the compression fronts travelling into uncompressed matter. The higher
compression will
also cause the compressed system to occupy a smaller spatial volume.

Let us now study the expansion stage in the one-dimensional
scenario. For the qualitative arguments
to be presented, it is sufficient to
restrict considerations to baryon-free matter and, since in this
case hadronic matter consists predominantly of (thermal) pions,
to a massless pion gas for the hadronic part of the EoS. This is a good
approximation for temperatures below $200$ MeV. Above that value, nucleons
become massless in the $\sigma-\omega$--model \cite{theis} and
contribute roughly the same amount to thermodynamic quantities as pions.
For our choice of the Bag constant, however, the phase transition to
the QGP happens prior to this phenomenon ($T_c \simeq 170$ MeV).

Employing this EoS for baryon-free matter,
we solve the hydrodynamic equations (\ref{dt}, \ref{dn})
with the relativistic HLLE algorithm \cite{test1}. The initial condition
is a blob of size $2\, R$ and constant energy density $\epsilon_0$.
In Fig.\ 4 we show the lifetime (in the CM frame)
of the mixed phase in the center of the compressed system
as a function of $\epsilon_0$ \cite{test1}.
One observes a pronounced peak around the ``softest point'' of the EoS
(near the energy density $\epsilon_Q$, i.e., at
the boundary between mixed phase matter and the QGP). How
this peak is related to the hydrodynamic expansion solution was
explained in detail in Ref.\ \cite{test1}. The reason for this
prolongation of the lifetime is that the system does not expand (and cool)
as rapidly as in the case without a phase transition, but stays
in a comparatively small spatial volume for a long time. The rapid cooling for
the case without phase transition
is illustrated by the dashed line which shows the CM time
when the temperature
of the {\em pure\/} pion gas drops below the temperature $T_c \simeq 170$
MeV in the center of the system\footnote{The notion of a ``phase
transition temperature'' is, of course, irrelevant in this case.}.

To conclude our investigation of the simple one-dimensional scenario,
the presence of a transition from hadron to quark and gluon degrees of freedom
in the nuclear matter EoS leads (a) to a prolonged compression stage where
the final values for energy and baryon number density are larger than
for the case without phase transition. Consequently, the zone of compressed
matter occupies a smaller spatial volume. It furthermore leads to
(b) a prolonged expansion stage, i.e., matter will not
expand and cool rapidly but stay in that relatively
small spatial volume for a long time.

Let us now turn to 3+1--dimensional hydrodynamical calculations. We first
consider a
Au+Au--collision at 5 AGeV (i.e., close to the ``softest'' point in the
EoS, cf.\ Fig.\ 2(e)) and finite impact parameter $b=3$ fm. For all
multi-dimensional calculations presented here we
use a SHASTA algorithm with first order accuracy in time \cite{test1,SHASTA}.
The grid spacing is chosen as $\Delta x = 0.3$ fm, the time step width
is $\Delta t = 0.4\, \Delta x$. The use of this rather coarse grid spacing
\cite{test1,test2} and the first order scheme (which is not as accurate
as a second order scheme \cite{test1}) is a concession to the enormous
calculational effort of 3+1--dimensional hydrodynamical calculations. For
the results presented below, however, we do not expect major quantitative
changes when using a finer grid and a second order scheme.

In Fig.\ 5 we show CM frame baryon density contours (and flow
velocity vectors) in the reaction plane at different CM times.
Part (a) employs the EoS with phase transition and part (b) the
pure hadronic EoS. One notices the following distinct feature:
for the EoS with phase transition the compressed zone in the
center of the reaction grows
and expands much more slowly (and also the compression is much higher) than
in the collision calculated with the pure hadronic EoS. Since this
is completely analogous to our findings in the one-dimensional scenario,
the previous considerations immediately present
the obvious explanation for this phenomenon also in three space dimensions.
In the latter case, however, this
has the following further consequence for spectator matter: since
compressed matter does not expand rapidly in the case of the EoS with
phase transition, it does not exert pressure onto the spectators,
which consequently pass the participants undeflected.
In contrast, for a pure hadronic EoS the compressed zone expands violently
and deflects spectator matter (the mentioned
``bounce--off'' effect \cite{stoecker}).

This effect can be seen best in the mean transverse (in--reaction--plane)
fluid momentum per baryon as a function of longitudinal fluid rapidity,
\begin{equation} \label{pxA}
\langle p_x/N \rangle (y) \equiv \frac{
\sum_i' N_{i}~p_{x,i}}{\sum_i' N_{i}}~.
\end{equation}
Here, the sum is over all fluid elements $i$ (i.e., in praxi
numerical grid cells) subject to the constraint that
their longitudinal rapidity $y_i \equiv \frac{1}{2}
\ln [(1+v_{z,i})/(1-v_{z,i})]$
obeys $y-\Delta y/2 \leq y_i \leq y+\Delta y/2$ ($v_{z,i}$ is
the $z$--component of the 3--velocity of fluid element $i$), and
$N_{i}$ is the total net baryon number contained in fluid element $i$.
Furthermore, $p_{x,i} \equiv M \gamma_i
v_{x,i}$ (here $\gamma_i= (1-{\bf v}_i^2)^{-1/2}$). We remark that
(\ref{pxA}) is not directly comparable to an experimentally measurable
quantity, since $\langle p_x/N \rangle (y)$
is essentially (a constant factor, $M$, times)
the average $x$--component of the {\em fluid\/} 4--velocity in the rapidity bin
$[y-\Delta y/2, y + \Delta y/2]$,
and {\em not\/} the $x$--component of the average nucleon momentum.
The latter can only be obtained after a suitable freeze--out procedure
is applied to the hydrodynamic quantities. This is the subject of a
forthcoming paper \cite{bernard}.

In Fig.\ 6 we show the quantity (\ref{pxA}) at the end of the collision
(defined as the CM time when the baryon number (fluid) rapidity distribution
does no longer change appreciably)
as a function of rapidity $y$ normalized to beam rapidity $y_{CM}$ for
Au+Au--collisions at $b=3$ fm and for beam energies
(a) $E_{Lab}^{kin}= 3.5$ AGeV, (b) $5$ AGeV, and (c) $11.7$ AGeV.
One observes that for the pure hadronic EoS (open circles) the
$\langle p_x/N\rangle (y)$--curves
have the familiar S--shape, representing transversally directed
collective flow of matter.
This matter corresponds to the deflected spectators of Fig.\ 5(b).
In all three cases the maximum of the transverse momentum is
slightly below beam rapidity (because the deflection of the spectators
decreases their longitudinal momentum) and of the order of $300$ MeV.
The maximum decreases slowly as a function of beam energy.

On the other hand, for the EoS with phase transition the transverse momentum
has an S--shaped form for low beam energies, Fig.\ 6(a),
although the maximum transverse momentum is by about a factor of two
smaller than for the case of a pure hadronic EoS. This is due
to the creation of a small amount of mixed phase matter in the very central
region at these energies (cf.\ Fig.\ 2)\footnote{We note that for central
cells the compression is near the limiting values given by the one-dimensional
collision scenario as shown in Fig.\ 2.}.
However, around $5$ AGeV (Fig.\ 6(b)) there is no longer any appreciable
amount of collective transverse flow, the transverse momentum
is essentially zero as function of rapidity. As explained in connection
with Fig.\ 5(a), spectators are not deflected at all in this case, they
pass the participant matter before the latter is expanding. This
expansion is then more or less isotropic.
For even higher beam energies, Fig.\ 6(c),
the directed flow of matter gradually starts to increase again, since
QGP with a higher $p/\epsilon$ is created (cf.\ Fig.\ 2(e)),
which does again expand (and cool) more rapidly, see Fig.\ 4, and
consequently deflects spectator matter.

In Fig.\ 7 we show the excitation function
of the {\em directed\/} transverse (in--reaction--plane) fluid momentum
per baryon,
\begin{equation}
\langle p_x/N \rangle^{dir} = \frac{1}{N} \int_{-y_{CM}}^{y_{CM}} {\rm d}y~
\langle p_x/N \rangle (y)~\frac{{\rm d}N}{{\rm d}y}~{\rm sgn} (y)~,
\end{equation}
which is in principle nothing else but the integral over the
curves of Fig.\ 6, weighted with the baryon number rapidity distribution
(and the sign of the rapidity, otherwise momentum conservation would yield a
trivial value). First of all, one observes that above
$E_{Lab}^{kin} \simeq 2$ AGeV this quantity decreases
for increasing beam energy for calculations with
the pure hadronic EoS (dotted line). This is in accord
with Figs.\ 6(a--c) and simply due to the fact that faster spectators
are less easily deflected by the hot, expanding participant matter.
Second, the directed transverse momentum as calculated
for the EoS with phase transition (full line) shows a dramatic {\em drop\/}
between BEVALAC and AGS beam energies
as compared to the calculation with the pure hadronic EoS and
{\em increases\/} again beyond $\sim 10$ AGeV.
Thus, {\em there is a local minimum in the excitation function of the
directed transverse (in--reaction--plane) collective flow around $\sim 6$\/}
AGeV, which is related to the phase transition to the QGP
and the existence of a ``softest point'' in the nuclear matter EoS.
We note that
the position of the minimum strongly depends on the EoS (cf.\ discussion
of Fig.\ 2). It may easily shift to higher beam energies, if resonances are
included in the hadronic part of the EoS. Also, as was the case for Fig.\ 6,
absolute values for the directed momentum cannot be compared to
experimentally measured ones, since a freeze-out calculation
has not yet been performed. Such a quantitative comparison is
in any case not reasonable at this stage since viscosity effects are
neglected in the ideal hydrodynamic picture, which are known to
have a strong influence on flow \cite{schmidt}\footnote{How to implement
dissipative effects into relativistic hydrodynamics is
a presently unsolved problem, see the discussion in \cite{strottman,st2}.}.

The main point is, however, that irrespective of these quantitative
uncertainties, the {\em minimum\/} is a clean {\em qualitative\/}
signal for a transition from hadron
to quark and gluon degrees of freedom in the nuclear matter EoS.
We even expect
this signal to be independent on whether the transition is a first order phase
transition or merely a rapid increase of the energy density as a function
of temperature (or, for baryon-rich matter, of the baryo-chemical
potential). Let us emphasize
that the mere fact that the flow {\em decreases\/} when
using an EoS with phase transition relative to the pure hadronic case,
as was found in \cite{bravina}, is not sufficient to serve as
an unambiguous QGP signature.
First of all, absolute values for the flow are not reliable due
to the above mentioned uncertainties in the ideal hydrodynamic picture.
Second, as can be seen in Fig.\ 7
such a decrease occurs also for the pure hadronic
scenario and is due to trivial kinematic reasons.
Finally, it was
suggested in \cite{ART} that an observed flow which is less than
predicted by cascade models might be a signature for QGP formation.
This, however, appears to be insufficiently unique as well:
as was shown in \cite{ARC,frankel}
there is considerable freedom in treating
two--particle scatterings in a cascade which results in quite
different values for the flow.

In order to observe the minimum in $\langle p_x/N \rangle^{dir}
(E_{Lab}^{kin})$ experimentally, it is mandatory
to supplement present data on collective flow
for BEVALAC \cite{rai} and AGS energies
\cite{Yingchao} with data taken
at beam energies between BEVALAC and present AGS energies, in order to map out
the excitation function of the transverse collective flow.
Such experiments are currently under way at the AGS \cite{proposal} and will
help to decide whether physics at these energies
is satisfactorily described by hadronic interactions, like
in microscopic models \cite{RQMD,ARC,ART}, or whether one has already entered
the rather interesting domain of qualitatively new phenomena that
can only be attributed to the presence of quark and gluon degrees of freedom.
We conclude by remarking that there are no data that
conclusively exclude the latter possibility. Rather,
the measured baryon rapidity distribution at AGS energies \cite{gonin}
can be equally well described by microscopic \cite{ARC}
and fluid-dynamical models \cite{bravina}.
\\~~\\
\noindent
{\bf Acknowledgements}
\\ ~~ \\
D.H.R.\ thanks Brian Cole, Miklos Gyulassy, Gulshan Rai,
Edvard Shuryak, and Bill Zajc for discussions
and Miklos Gyulassy for his continuous interest and encouragement
that contributed essentially to the completion of this work.
\\~~\\

\newpage
\noindent
{\bf Figure Captions:}
\\ ~~ \\
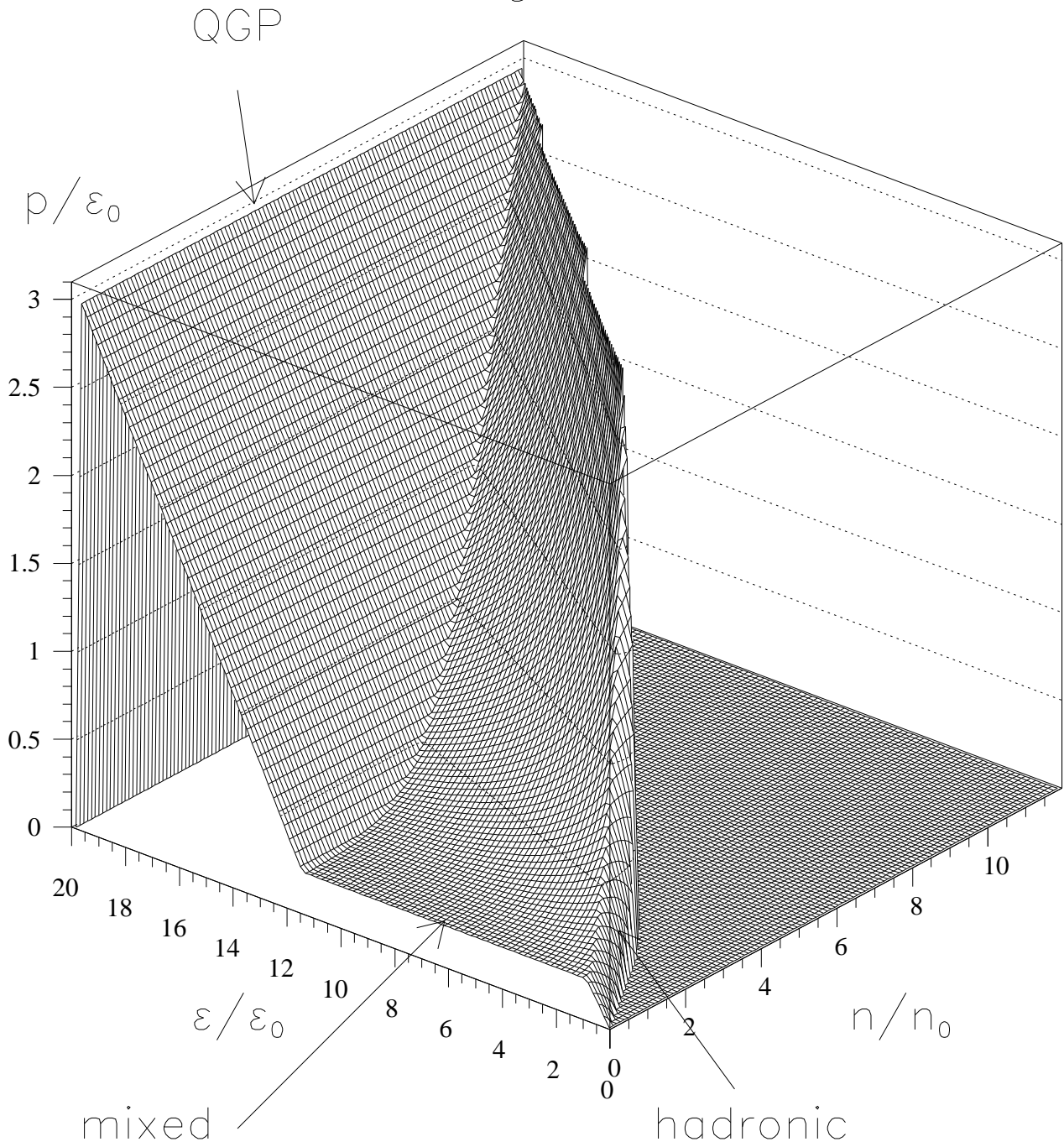
{\bf Fig.\ 1:} The nuclear matter EoS in the form
$p(\epsilon,n)$. Pressure and energy density are normalized to the
ground state energy density $\epsilon_0 \simeq 146.5\, {\rm MeV\, fm}^{-3}$,
the baryon number density to the ground state density $n_0 \simeq 0.16\,
{\rm fm}^{-3}$.
\\ ~~ \\
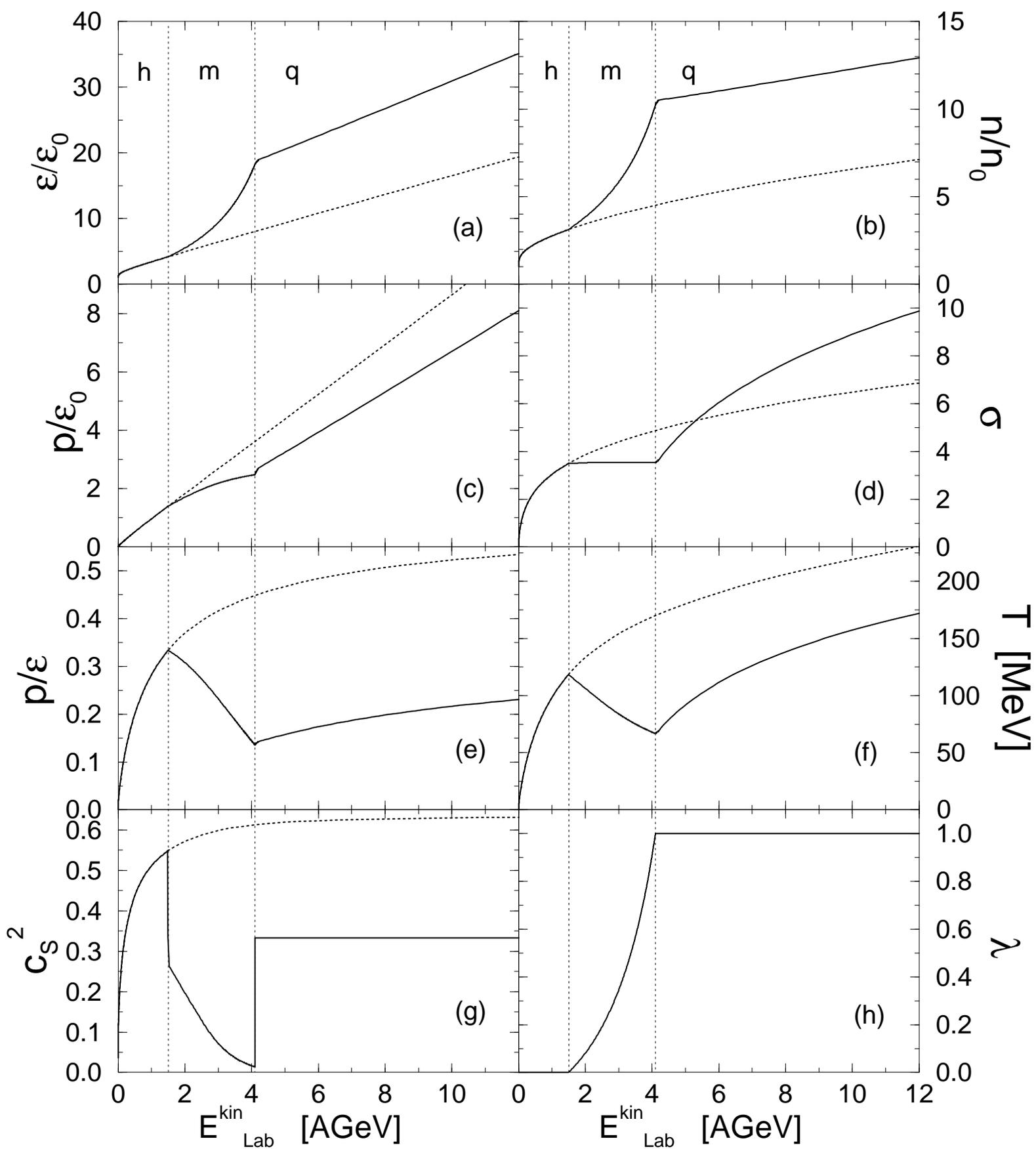
{\bf Fig.\ 2:} Excitation functions of various thermodynamic
quantities for the one-dimensional ``slab-on-slab'' collision scenario.
(a) $\epsilon/\epsilon_0$, (b) $n/n_0$, (c) $p/\epsilon_0$, (d) $\sigma$,
(e) $p/\epsilon$, (f) $T$, (g) the velocity of sound squared $c_s^2$, and (h)
the volume fraction of QGP $\lambda$ in the final compressed state.
Full lines are for the EoS with
phase transition, dotted lines for the pure hadronic EoS. Regions of
pure hadronic, mixed, and pure quark matter in the final compressed
state are indicated by thin dotted lines.
\\ ~~ \\
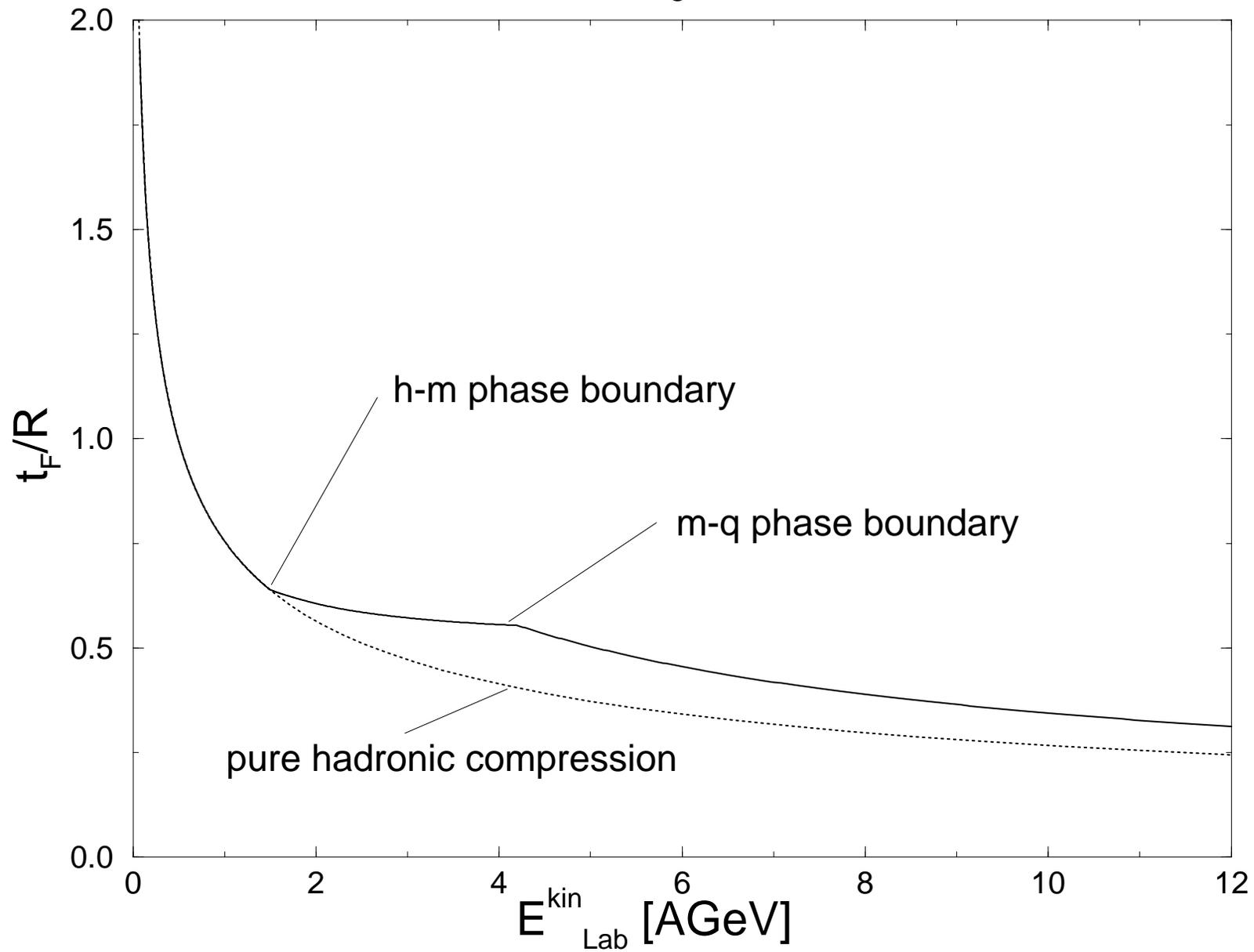
{\bf Fig.\ 3:} The time $t_F$ a compression front requires to
traverse an incoming nucleus in units of the nuclear rest frame
radius $R$ as a function of the beam energy. The full line is for
the EoS with phase transition, the dotted for the pure hadronic EoS.
\\ ~~ \\
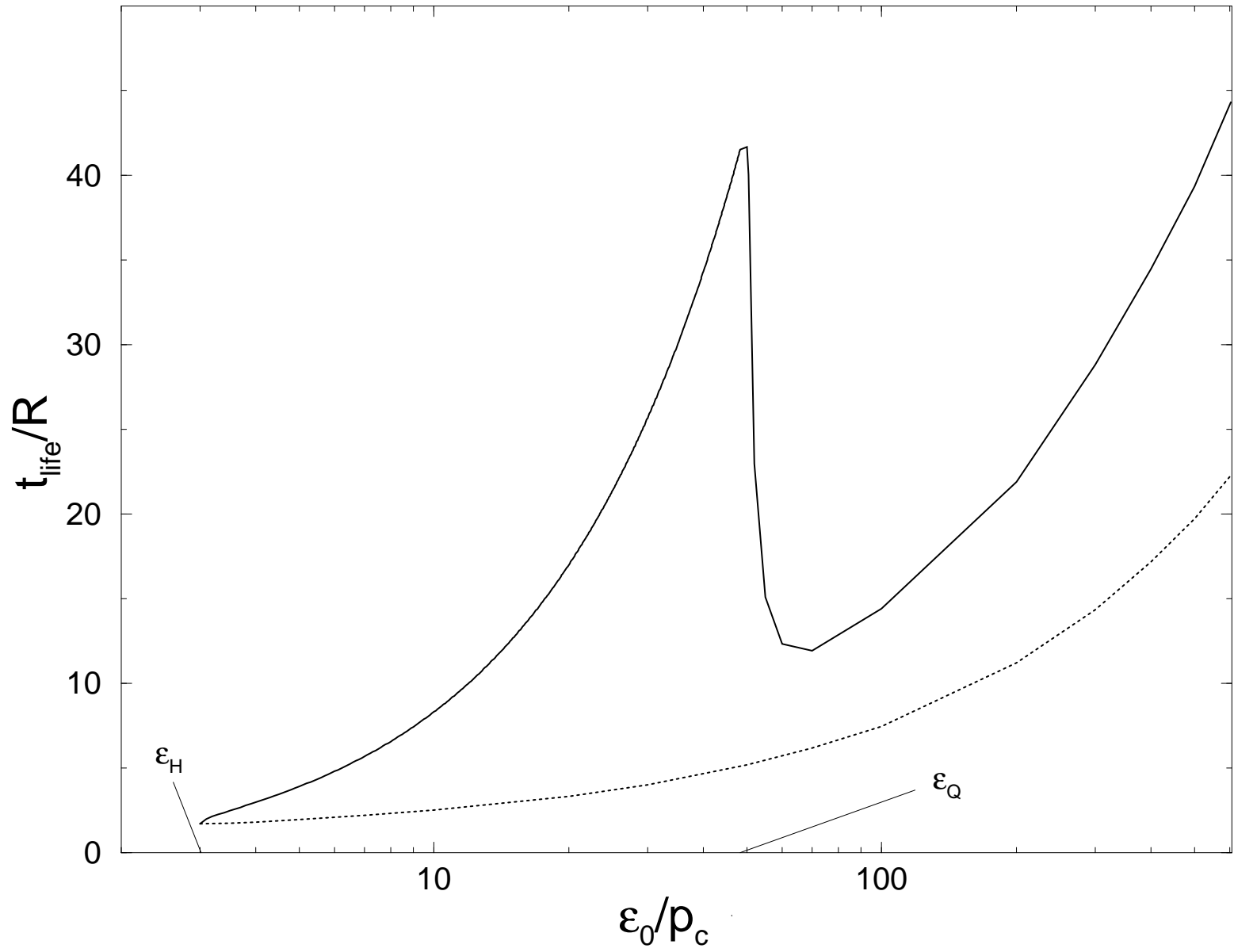
{\bf Fig.\ 4:} The lifetime of the mixed phase, defined as the CM time
when the $T_c$--isotherm intersects the $t$--axis at $x=0$, in units
of the radius of the hot, compressed system as a function of the
initial energy density (in units of the phase transition pressure $p_c$).
$\epsilon_H$ and $\epsilon_Q$ indicate the phase boundaries between
hadron matter and mixed phase as well as mixed phase and QGP.
The full line is for the EoS with phase transition, the dotted line
for the pure hadronic EoS, assuming $T_c \simeq 170$ MeV.
\\ ~~ \\
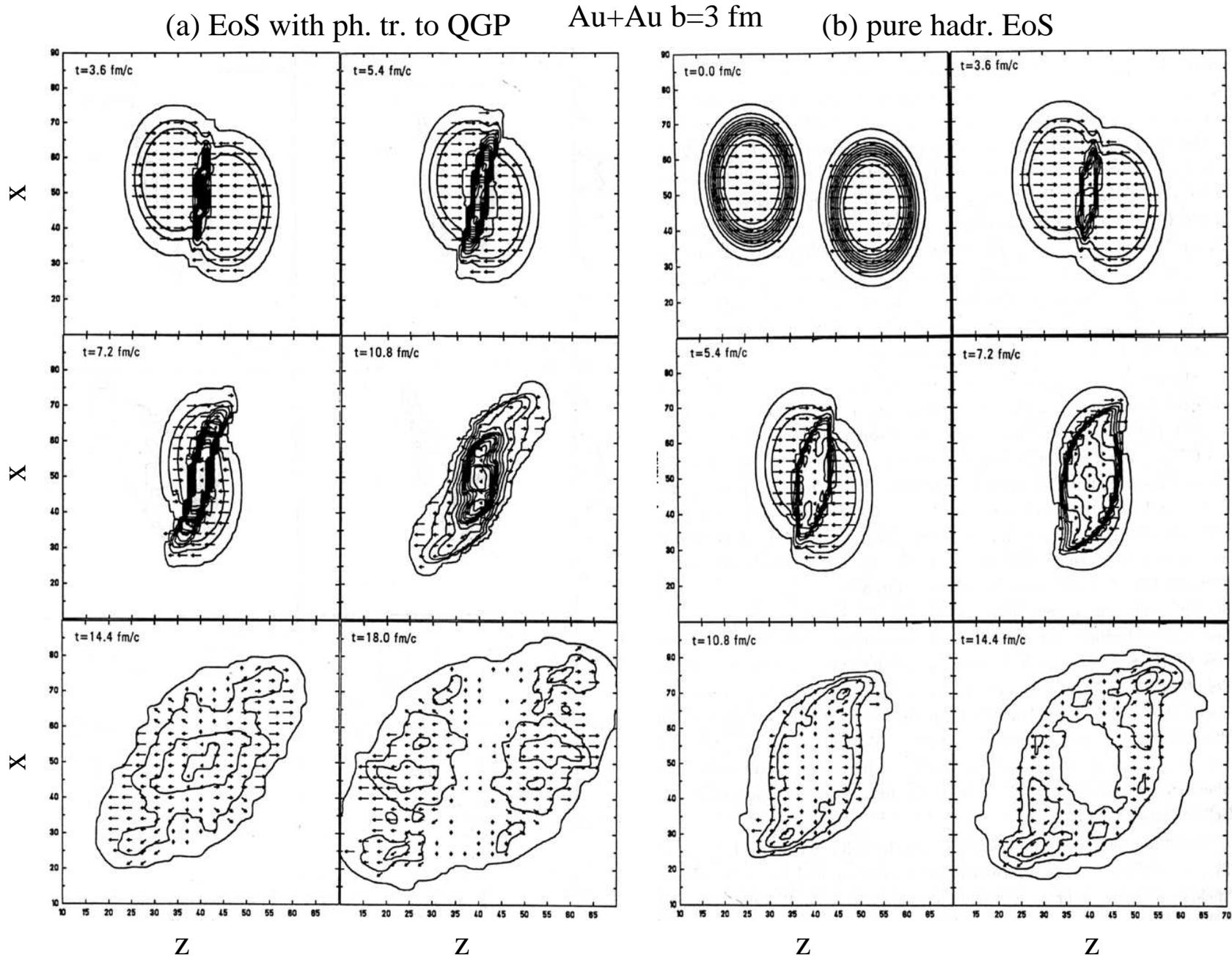
{\bf Fig.\ 5:} CM frame baryon density contours and flow velocity vectors
in the reaction plane ($x-z$--plane)
at different CM times for a $5$ AGeV Au+Au--collision at $b=3$ fm
calculated with (a) the EoS with phase transition and (b) the
pure hadronic EoS.
\\ ~~ \\
{\bf Fig.\ 6:} The mean transverse (in--reaction--plane) momentum per
baryon as a function of longitudinal (fluid) rapidity for Au+Au--collisions
at $b=3$ fm and for
(a) $E_{Lab}^{kin} = 3.5$ AGeV, (b) $5$ AGeV, and (c) $11.7$ AGeV.
The crosses correspond to calculations using the EoS with phase transition,
the open circles to those with the pure hadronic EoS.
Full and dash--dotted lines
are interpolations between calculated points to guide the eye.
\\ ~~ \\
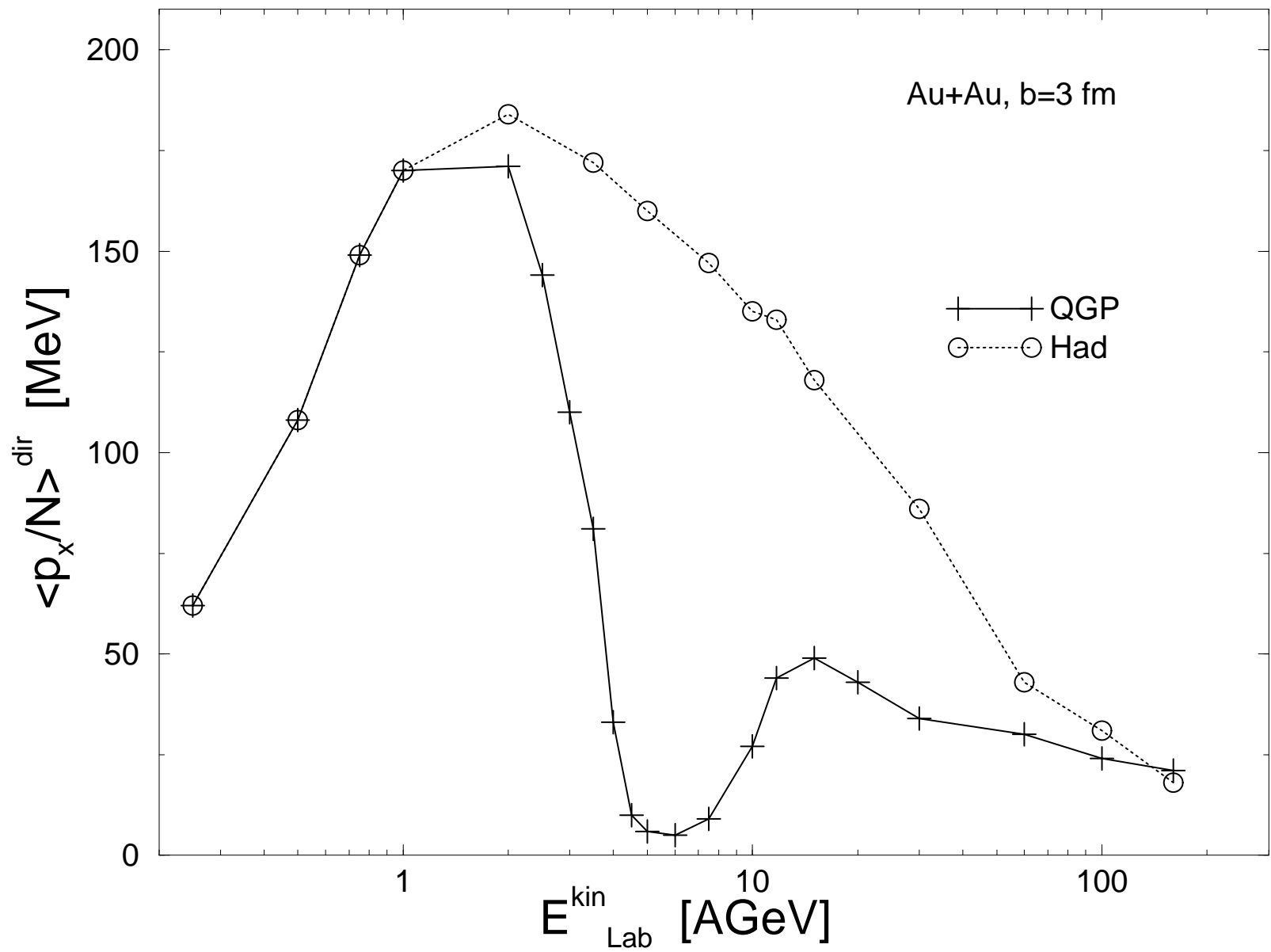
{\bf Fig.\ 7:} The directed mean transverse (in--reaction--plane)
momentum as a function of beam energy for Au+Au--collisions at $b=3$ fm.
The full line (crosses) corresponds to calculations using the EoS
with phase transition, the dotted line (open circles) to those with the
pure hadronic EoS.

\begin{thebibliography}{99}
\bibitem{maruhn} J.A.\ Maruhn, W.\ Greiner, in: ``Treatise on Heavy--Ion
Science'', Vol.\ 4 (ed.\ D.A.\ Bromley, Plenum Press, New York, London, 1985),
p.\ 565.
\bibitem{stoecker} H.\ St\"ocker, W.\ Greiner, Phys.\ Rep.\ 137 (1986) 277.
\bibitem{strottman} R.B.\ Clare, D.D.\ Strottman, Phys.\ Rep.\ 141 (1986) 177.
\bibitem{st2} D.\ Strottman, Nucl.\ Phys.\ A 566 (1994) 245c.
\bibitem{gutbrod} H.A.\ Gustaffson et al., Phys.\ Rev.\ Lett.\ 52 (1984)
1590, \\
H.H.\ Gutbrod, K.H.\ Kampert, B.W.\ Kolb, A.M.\ Poskanzer, H.G.\ Ritter,
H.R.\ Schmidt, Phys.\ Lett.\ B 216 (1989) 267.
\bibitem{rai} H.G.\ Ritter and the EOS collaboration,
Nucl.\ Phys.\ A 583 (1995) 491c, \\
M.D.\ Partlan and the EOS collaboration, preprint LBL-36280, UC-414
(submitted to Phys.\ Rev.\ Lett.).
\bibitem{rentzsch}
U.\ Ornik, F.W.\ Pottag, R.M.\ Weiner, Phys.\ Rev.\ Lett.\ 63 (1989) 2641, \\
T.L.\ McAbee, J.R.\ Wilson, J.A.\ Zingman, C.T.\ Alonso, Mod.\ Phys.\ Lett.\
A 4 (1989) 983, \\
B.\ Waldhauser, D.H.\ Rischke, U.\ Katscher, J.A.\ Maruhn, H.\ St\"ocker,
W.\ Greiner, Z.\ Phys.\ C 54 (1992) 459.
\bibitem{csernai}
N.S.\ Amelin, E.F.\ Staubo, L.P.\ Csernai, V.D.\ Toneev, K.K.\ Gudima,
D.\ Strottman, Phys.\ Rev.\ Lett.\ 67 (1991) 1523, Phys.\ Lett.\ B 261
(1991) 352,\\
N.S.\ Amelin, L.P.\ Csernai, E.F.\ Staubo, D.\ Strottman, Nucl.\ Phys.\ A 544
(1992) 463c.
\bibitem{bravina} L.V.\ Bravina, N.S.\ Amelin, L.P.\ Csernai, P.\ Levai, D.\
Strottman, Nucl.\ Phys.\ A 566 (1994) 461c.
\bibitem{prakash} R.\ Venugopalan, M.\ Prakash, M.\ Kataja, P.V.\
Ruuskanen, Nucl.\ Phys.\ A 566 (1994) 473c.
\bibitem{QMD} for a review, see: G.\ Peilert, H.\ St\"ocker, W.\ Greiner, Rep.\
Prog.\ Phys.\ 57 (1994) 533.
\bibitem{RQMD} H.\ Sorge, H.\ St\"ocker, W.\ Greiner,
Ann.\ Phys.\ (N.Y.) 192 (1989) 266.
\bibitem{ARC} D.E.\ Kahana, D.\ Keane, Y.\ Pang, T.\ Schlagel, S.\ Wang,
preprint nucl-th/9405017,\\
S.H.\ Kahana, T.J.\ Schlagel, Y.\ Pang, Nucl.\ Phys.\ A 566 (1994) 465c.
\bibitem{ART} Bao-An Li, C.M.\ Ko, G.Q.\ Li, preprint nucl-th/9502047.
\bibitem{QM93} see e.g.:
Proc.\ of ``Quark Matter '93'', Nucl.\ Phys.\ A 566 (1994).
\bibitem{QGP} see e.g.: B.\ M\"uller, ``The physics of the quark--gluon
plasma'', Lecture Notes in Physics, vol.\ 225, (Springer, New York, 1985).
\bibitem{bubble} L.P.\ Csernai, J.I.\ Kapusta, Phys.\ Rev.\ Lett.\ 69
(1992) 737.
\bibitem{werner} K.\ Werner, Nucl.\ Phys.\ A 566 (1994) 477c.
\bibitem{test1} D.H.\ Rischke, S.\ Bernard, J.A.\ Maruhn,
preprint CU-TP-692, nucl-th/9504018 (submitted to Nucl.\ Phys.\ A).
\bibitem{test2} D.H.\ Rischke, Y.\ P\"urs\"un, J.A.\ Maruhn,
preprint CU-TP-693, nucl-th/9504021 (submitted to Nucl.\ Phys.\ A).
\bibitem{bugaev} K.A.\ Bugaev, M.I.\ Gorenstein, B.\ K\"ampfer, V.I.\ Zhdanov,
Phys.\ Rev.\ D 40 (1989) 2903.
\bibitem{olli} J.P.\ Blaizot, J.Y.\ Ollitrault, Phys.\ Rev.\ D 36 (1987) 916.
\bibitem{DHRMG} D.H.\ Rischke, M.\ Gyulassy (in preparation).
\bibitem{hung} C.M.\ Hung, E.V.\ Shuryak, preprint SUNY-NTG-94-59,
hep-ph/9412360.
\bibitem{hofmann} J.\ Hofmann, H.\ St\"ocker, U.\ Heinz, W.\ Scheid, W.\
Greiner, Phys.\ Rev.\ Lett.\ 36 (1976) 88.
\bibitem{LL6} L.D.\ Landau, E.M.\ Lifshitz, ``Fluid mechanics'' (Pergamon,
New York, 1959).
\bibitem{MIT} A.\ Chodos, R.L.\ Jaffe, K.\ Johnson, C.B.\ Thorn,
V.F.\ Weisskopf, Phys.\ Rev.\ D 9 (1974) 3471.
\bibitem{migdhr} M.I.\ Gorenstein, D.H.\ Rischke, H.\ St\"ocker, W.\ Greiner,
J.\ Phys.\ G 19 (1993) L69.
\bibitem{walecka} see e.g.: B.D.\ Serot, J.D.\ Walecka,
``The Relativistic Nuclear Many--Body Problem'' in:
Adv.\ Nucl.\ Phys.\ 16 (1986) 1 (eds.\ J.W.\ Negele and E.\ Vogt, Plenum Press,
New York).
\bibitem{BGR} K.A.\ Bugaev, M.I.\ Gorenstein, D.H.\ Rischke,
Phys.\ Lett.\ B 255 (1991) 18.
\bibitem{RiFri} D.H.\ Rischke, B.L.\ Friman, H.\ St\"ocker, W.\ Greiner,
J.\ Phys.\ G 14 (1988) 191.
\bibitem{theis} J.\ Theis, G.\ Graebner, G.\ Buchwald, J.\ Maruhn, W.\ Greiner,
H.\ St\"ocker, J.\ Polonyi, Phys.\ Rev.\ D 28 (1983) 2286.
\bibitem{SHASTA} J.P.\ Boris, D.L.\ Book, J.\ Comput.\ Phys.\ 11 (1973) 38,\\
D.L.\ Book, J.P.\ Boris, K.\ Hain, J.\ Comput.\ Phys.\ 18
(1975) 248.
\bibitem{bernard} S.\ Bernard, J.A.\ Maruhn, W.\ Greiner, D.H.\ Rischke
(in preparation).
\bibitem{schmidt} W.\ Schmidt, U.\ Katscher, B.\ Waldhauser, J.A.\ Maruhn,
H.\ St\"ocker, W.\ Greiner, Phys.\ Rev.\ C 47 (1993) 2782.
\bibitem{frankel} M.\ Gyulassy, K.A.\ Frankel, H.\ St\"ocker, Phys.\ Lett.\
B 110 (1982) 185.
\bibitem{Yingchao} J.\ Barrette et al.\ (E877 collaboration),
Phys.\ Rev.\ Lett. 73 (1994) 2532, and Proc.\ of ``Quark Matter '95''
(to appear in Nucl.\ Phys.\ A),\\
Y.\ Zhang and J.P.\ Wessels for the E877 collaboration,
Proc.\ of ``Quark Matter '95'' (to appear in Nucl.\ Phys.\ A).
\bibitem{proposal} G.\ Rai and the E895 collaboration, LBL PUB--5399 (1993).
\bibitem{gonin} M.\ Gonin for the E802/E866 collaboration,
Nucl.\ Phys.\ A 553 (1993) 799c.
\end{thebibliography}
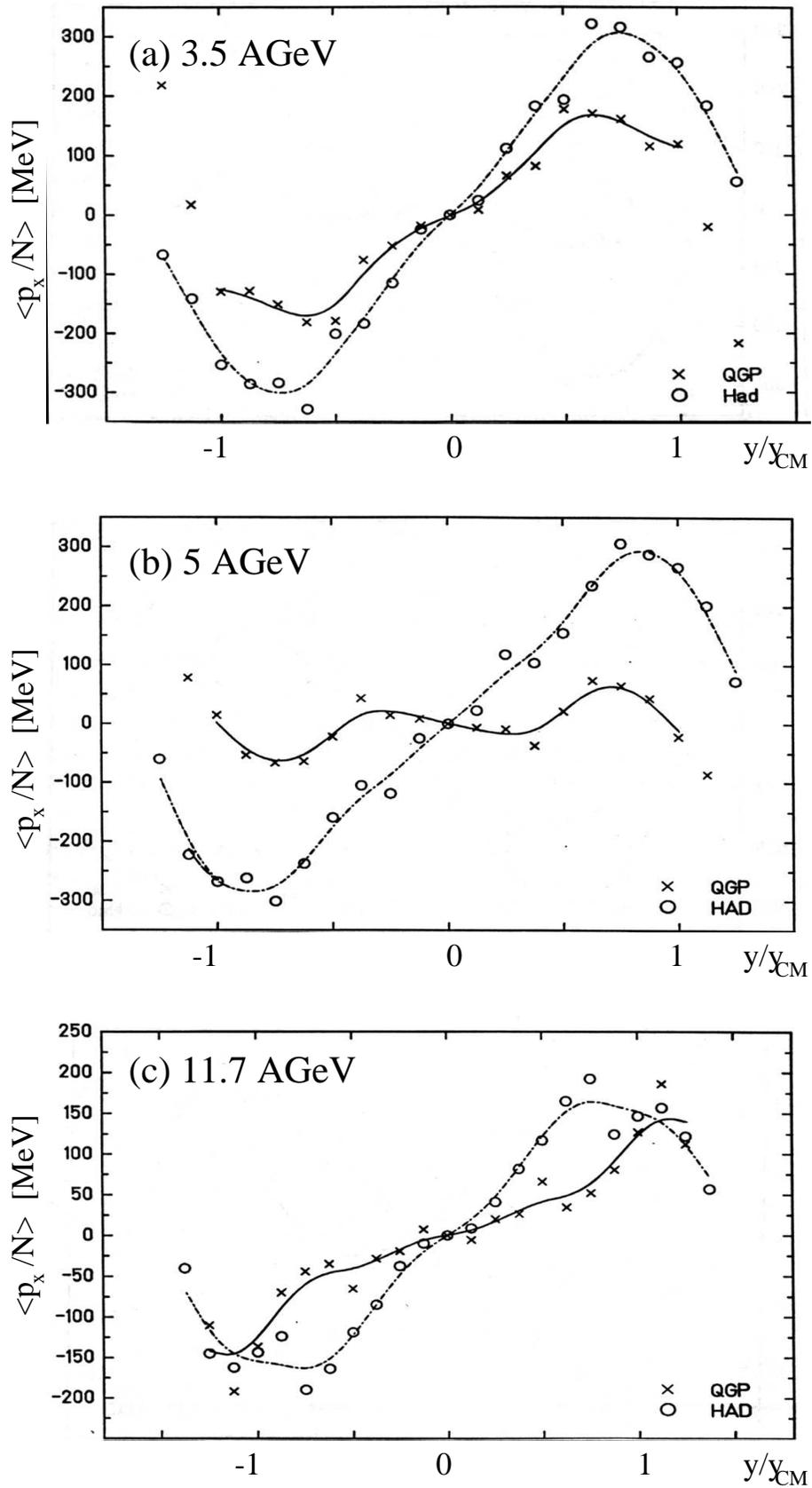
\end{document}